# Chapter 1
# Social Big Data: An Overview and Applications

Bilal Abu-Salih[1], Pornpit Wongthongtham[2]
Dengya Zhu[3] , Kit Yan Chan[3] , Amit Rudra[3]

[1]The University of Jordan
[2] The University of Western Australia
[3] Curtin University

**Abstract :** The emergence of online social media services has made a qualitative leap and brought profound changes to various aspects of human, cultural, intellectual, and social life. These significant Big data tributaries have further transformed the businesses processes by establishing convergent and transparent dialogues between businesses and their customers. Therefore, analysing the flow of social data content is necessary in order to enhance business practices, to augment brand awareness, to develop insights on target markets, to detect and identify positive and negative customer sentiments, etc., thereby achieving the hoped-for added value. This chapter presents an overview of Social Big Data term and definition. This chapter also lays the foundation for several applications and analytics that are broadly discussed in this book.

**Keywords:** Social Big Data; Social Credibility; Domain Knowledge; Sentiment Analysis; Affective Design; Predictive Analytics;

## 1.1   Introduction

The social media services, positioned on the throne of cyberspace, in their broad sense, cover an ample set of freely accessible electronic platforms that are built to encourage and simplify communication between people with similar interests by enabling interactive conversations and exchanging information regardless of physical location. Those virtual platforms are continuing to spread exponentially by providing social communication services to their affiliated members. The services offered by these sites have expanded, providing their consumers with extensive possibilities for exchanging information in the fields of education, health, culture, sports and other domains of knowledge [1, 2].

In modern business firms, social media services are incorporated as part of the infrastructure for a number of emerging applications such as personalized recommendation systems [3, 4], opinion analysis [5], expertise retrieval [6, 7], and computational advertising [8, 9]. In such applications, social data offers a plethora of benefits to enhance the decision making process. Business intelligence applications are more focused on structured data; however, in order to understand and analyse the social media data, there is a need to aggregate data from various sources and to present it in a plausible format. Hence, "many marketing researchers believe that social media analytics presents a unique opportunity for businesses to treat the market as a 'conversation' between businesses and customers" [10]. Social Big Data (SBD) [11] exhibit all the typical properties of big data: wide physical distribution, diversity of formats, non-standard data models, independently-managed and heterogeneous semantics.

In this context, social data analysis is an evolving task and join various disciplines such as social media analysis, semantic discovery, predictive analytics, sentiment analysis, affective design and big data computing [12-25]. For example, as the SBD are derived from a variety of sources, it is essential to measure the reputation of the source and provide flexibility to the analysts, so that the trust value of each source can be understood [26]. Another important reflection is the semantics of extracted textual data from which meaningful information can be derived. Also, developing opinion mining and sentiment analysis techniques to extract and summarise sentiment data effectively can assist to hear Voice of the Customer (VoC) [27] and Voice of the Market (VoM) [28] from social media. Last, but not least, the era of social big data has exposed several fertile resources to discover and collect large scale of big affective data. Therefore, the trusted and meaningful external data that cover the global environment, the VoM, and the VoC, can be collected and stored for further analysis. However, due to the massive amount of information produced by these platforms, in conjunction with the absence of a gatekeeper for those sites, it is difficult to verify the credibility of content and users. Therefore, the online social services are hijacked, and their valuable tools are used to spread chaos and misinformation. Hence, it is indispensable to have an accurate understanding of the contextual content of social users and their content, in order to establish a ground for measuring their social credibility consequently. Further, it is important to classify users and their content into appropriate categories prior to undertaking further business analytics.

This chapter presents a brief introduction to this book; first, an overview of the notion of Social Big Data is given followed by introducing various types of social data services as well as the importance and challenges of the exponentially increasing social data. Second, an array of substantial applications in the era of social big dats is discussed, this includes (i) the motivation for an approach to address the social big data problem is particularised by demonstrating the importance of determining the domains of interest of users and their content which leads to improving the forecasting of their future interest(s). (ii) The significance of deriving knowledge and measuring the credibility of the content of the online social platforms are discussed. (iii) A discussion is given on how social big data can be used to perform affective design of new products, which satisfy the product affective needs and aesthetic appreciation of developing new products.

## 1.2 SBD: An Overview

Since the advent and proliferation of Web 2.0, the role of web browsers has changed to enable users to send and receive content by means of several online tools that commenced with e-mail applications, chat, and chat forums that evolved into more recent and revolutionary electronic platforms such as social networks. These platforms provide an important means by which communities can grow and consolidate, allowing individuals or groups to share concepts and visions with others. Moreover, in addition to playing an active and distinctive role as effective media of social interaction, these social networks allow users to become acquainted with and understand the cultures of different peoples [29].

This rapid growth of the provided online social services and the explosive evolution of social data have established new research venues and produced new dissimilar notions to help comprehending the social impact of such digital environment. Hence, Social Big Data (SBD) and Big Social Data (BSD) notions have manifested as a combination of two terms – social media and Big Data – and are used interchangeably -and in this book as well- in reference to the massive amount of user-generated content, mainly in the form of unstructured data such as posts, photos, audios, videos etc.

### 1.2.1 Definition of SBD

There are few attempts to provide a formal definition to the term of SBD. This concept is defined by Bello-Orgaz et al. [11] as:

> *"Those processes and methods that are designed to provide sensitive and relevant knowledge to any user or company from social media data sources when data sources can be characterised by their different formats and contents, their very large size, and the online or streamed generation of information."*

Another attempt to provide a meta-level definition of the synthesized BSD concept is given by Olshannikova et al. [30] as:

> *"Big Social Data is any high-volume, high-velocity, high-variety and/or highly semantic data that is generated from technology-mediated social interactions and actions in digital realm, and which can be collected and analyzed to model social interactions and behavior."*

SBD was also identified as a resultant interdependence between the physical world and the social virtual world. Hence, Hiroshi Ishikawa [31] portrayed the SBD as a science of:

> *"analyzing both physical real world data (heterogeneous data with implicit semantics such as science data, event data, and transportation data) and social data (social media data with explicit semantics) by relating them to each other"*.

We can draw from the above definitions that SBD can be characterized with the same commonly features used to describe the notion of Big data. SBD is related to Big data paradigm in essence that it requires the same technology and sophisticated tools to analyse it. Therefore, SBD is a primal Big data island provides a momentum dense of social data which require a deep scrutinising.

SBD is perceived as a combination of three interrelated aspects, namely *contents* generated from social media, *infrastructure* to handle the high volume and speed of the propagated contents, and *analytics* to gain valuable insights. Discussions on infrastructure and analytics are extensively elaborated in the next chapters. In the following subsection an overview of a selective array of different types of social media is provided.

### 1.2.2 Types of social data services

Social data in general can be generated from a set of web-enabled portals and applications that facilitates the process of creating, editing, disseminating various types of user-generated contents [32, 33]. The following are examples of these social services.

**Online Social Networks (OSNs):** OSNs such as Facebook®, Twitter®, LiveBoon®, Orkut®, Pinterest®, Vine®, Tumblr®, Google Plus®, Instagram® etc, are relevant sources of data for SBD which enable users to create, edit and share videos, photos, files and instant conversations. OSNs have thrown open the doors of platforms for people to unleash their opinions and build new varieties of social communications based on these virtual societies. The vast amount of social data has spread to many different areas in everyday life such as e-commerce [32], education [34], health [35], to name a few. For example, several modern computing applications such as online education, weight loss and public health, music and entertainment rely on the content generated by OSNs [36]. This is evident in the dramatic increase in the use of these platforms for networking and communication. The Pew Research Center reported that 70% of American adults in Nov 2016 used OSNs for social interactions compared to 5% usage by the same user category in 2005 [37]. In Australia, the statistics for OSNs usage in Jan 2017 indicated around 2.8 million Twitter active users, 14.8 million visits to YouTube, 4.0 million Snapchat active users [37]. Such a dramatic connectivity with online social platforms has established a common ground that brings together people with shared interests, ideas and goals. **Error! Reference source not found.** shows the most popular OSNs as in April 2020.

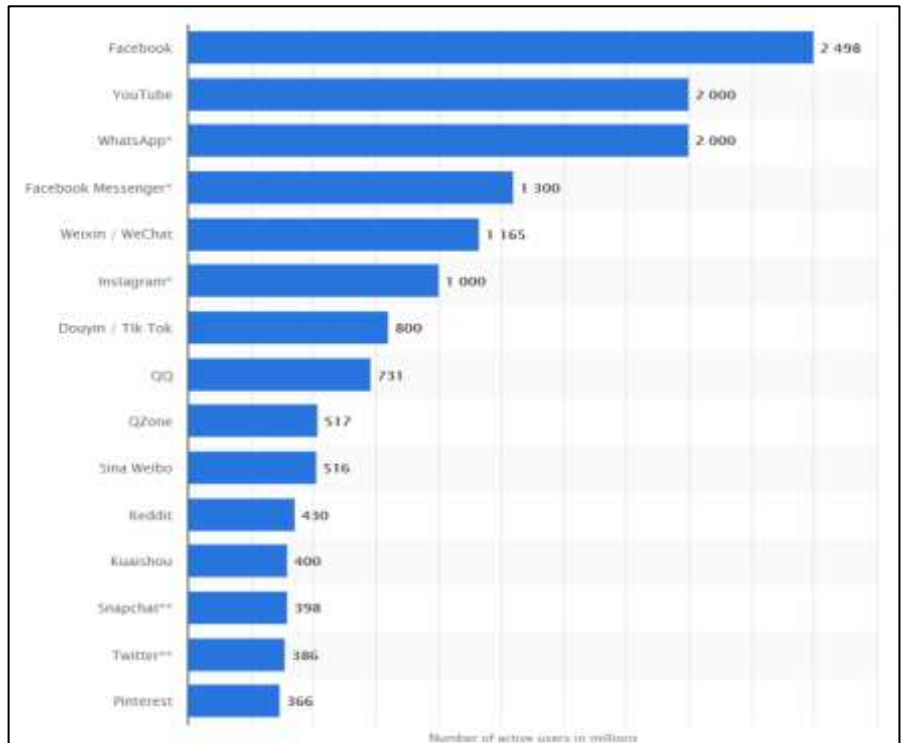

**Figure 0-1: Most popular social networks worldwide as of April 2020 [38]**

**Wikis:** Wiki is an online venue with an open-editing feature allows a person or a group of people to contribute to its content by allowing them adding and modifying content in numerous domains and topics. The mechanism provided by Wikis offers, for example, a space for team members to access and collaborate on a certain report simultaneously, thereby fortifying efforts to create, manage and disseminate knowledge and also working towards a common goal to benefit the organization [39]. In particular, Wikis are a good platform for businesses to conduct internal collaboration between staff members, it also provides an advanced system to work cooperatively on relevant internal documentations. Examples of Wikis include, but not limited to, Wikipedia, Wikitravel, WikiHow, WikiBooks, Wiktionary and Wikispecies.

**Social reviews websites:** review sites are online platforms offer users capacity to share their sentiments and opinions on certain products, services, businesses and even on people in terms of text, ratings, etc. Social reviews websites empower customers by providing them a Word-of-Mouth podium to unleash their opinions and recommendations on what they experienced. These social websites manage the collected evaluations and reviews and present them in various visualised forms to allow users locating relative and recommended products and services that match their preferences. Poor recommendations can negatively affect the company's reputation yet offer them a room for enhancement. Good reviews, on the other hand, elevate the company's position and improve visibility and their brand awareness [40-42]. Examples of social reviews websites include Google, Amazon, Yelp, TripAdvisor, YelloPages, etc.

**Discussion forums:** discussion forums are online sites provide the user the ability to discuss and share their thoughts, feelings, desires and opinions asynchronously [43]. They are identified as the earliest form of the currently well-known OSNs, yet provide distinguished platform for people to debate using a predefined set of categories, commonly divided in topics [44]. Also, these interactive online websites allow people to ask specific questions or inquiries and provide the opportunity for others to answer them in threaded conversational sequences [45]. Discussion forums can be used in different contexts; for example, educational providers can find the architecture of their electronic bulletin board feasible to conduct teaching practices using these platforms. This has proven success in various educational aspects [46].

**Video hosting and sharing sites:** video hosting and sharing are these online mediums facilitate uploading, downloading and disseminating personal and business related videos [47] . They provide companies, educational institutions, community organisations, individuals, and several categories of the societies an alternative media where they can store, manage, edit, and conduct analytics on their videos. Further, various video-hosting websites offer users the ability to commercialise their videos, add restrictions and policies on their contents, integrate with other social media sites, collaborate with other team members and share videos internally, allow audience to comment on videos and give feedback, to other assorted benefits. Examples of video sharing websites are; YouTube, Vimeo, Jetpack Video, Wistia, Vidyard, to name a few.

**Weblogs:** a weblog or a blog is an online journal website allows content to be displayed in a reverse chronological order, where the recent posted content appears first [48]. These sites are commonly managed by individuals(e.g. personal journal/diary) or a small group of people(e.g. community blog/ small business) and embody information pertains to specific or various topics, stories, events, etc. Interactivity is present in these websites; visitors are usually permitted to comment and give feedback on published entries. The set of technical skills and resources required to publish, manage and share contents on bogs is called blogging [49]. A special form of blogs is vlog(video blog) which mainly incorporates YouTube to publish and broadcast embedded videos and multimedia contents. Wordpress, TypePad and Tumblr are examples of content management system platforms used to create and manage weblogs.

**Sharing economy networks**: the notion of "sharing economy" has emerged as a new economic phenomenon supported by advances in information and communication technologies [50]. The term is defined by The Oxford English Dictionary (OED) as "an economic system in which assets or services are shared between private individuals, either for free or for a fee, typically by means of the Internet".  This peer-to-peer online model spans to various industries including travel, hotels, car sharing, staffing, as well as music and video streaming, and has brought tremendous benefits to consumers [51]. An example of sharing economy network is Airbnb, allows hosts to rent their properties, or spare rooms in their houses to anonymous guests with relatively cheaper prices than traditional hotels.

## 1.3 SBD applications and analytics

The challenge of managing and extracting useful knowledge from social media data sources has attracted much attention from academia and industry. This vast spread of social data necessitate researchers to obtain a better understanding of the massive amount of data being generated every second, leveraging of new data analysis techniques and the continuous improvement of existing practices. This section opens a on a preliminary dialogue on certain aspects, applications and analytics that are broadly discussed in this book.

### 1.3.1 Credibility of SBD

The changing role of online users from information consumers to information producers has caused a noticeable variance in the quality of published content [26]. In fact, quality of content is considered as a key difference between the content generated before and after the revolution of the Web 2.0 [52]. In this context, OSNs have been extensively used as a powerful tool to promote diffusion of information in several domains [53-56]. Given such an impact, an understanding and comprehension of the content of OSNs has been an essential interest of various research avenues [57]. In particular, identifying, reviewing, inferring and interpreting reputable social content consume a significant amount of time and effort [58], yet have attracted wide interest due to the significance of obtaining and applying high quality content in many disciplines such as politics [59], e-commerce [60], e-learning [61], and health care [62].

As discussed previously, data are no longer generated only by transactional/structured and limited external sources; the global environment is now producing data in the form of news, economic factors etc., and VoM and VoC through social networks, web blogs, etc. However, all external data sources do not have the same level of reputation. Data-users rely on reliable, reputable, and high-quality data and data sources. Likewise, unreliable and/or inaccurate data, such as data generated by suspicious and untrustworthy sources negatively impact on a company's operations and the decision making [63].

The quality of the data, which depends on whether it is collected from a reputable or an untrustworthy source, affects the quality of the perceived knowledge. For instance, in dramatic natural disasters such as the earthquake in Haiti and the tsunami in Japan, people used OSNs to report injury, share urgent and vital information, report damage, and provide firsthand observations [64-68]. However, while OSNs provide platforms for legitimate and genuine users, they also enable spammers and other untrustworthy users to publish and spread their content, taking advantage of the open environment and fewer restrictions which these platforms facilitate. This might lead some users to abuse OSNs platforms and hijack events such as emergency situations by spreading rumours, and false and misleading information [69]. Hence, studying users' behaviour in OSNs will lead to a better understanding of their published content. The users' behaviour comprises several social activities such as establishing new friendships, posting new content or replying to another user's content, messaging, browsing and discovery [70]. Furthermore, an analysis of the users' behaviour helps to determine and understand users' main topic(s) of interest [71], to mine their sentiments [72], and to know their needs and demands.

### 1.3.2 Domain of interest in SBD

Many individuals use OSNs to seek and connect with like-minded people. This homophily results in building homogenous personal networks in term of behaviours, interests, feelings, etc. [73]. In particular, OSNs provide a medium for content makers to express and share their thoughts, beliefs, and domains of interest. This gives individuals access to a wider audience which positively affects their social status and would assist them to obtain, for instance, political support [74]. Therefore, the cornerstone of the users' online social profiles is an accurate understanding of their domains of interest.

The domain of knowledge is a particular area of people's work, expertise, or specialisation within the scope of subject-matter knowledge (e.g. Sports, Politics, Information Technology, Education, Art and Entertainments, etc.) [75]. In online social services, the domains of interest can be determined at the user level and at the post level. In other words, the overall published content of the user is analysed, and the domain(s) of interest is inferred. Likewise, the user's posts can be analysed separately to extract the domain(s) of each post. The factual grasp of the users' domain(s) of interest facilitates understanding the domain(s) conveyed from a short text message such as a *tweet*.

### 1.3.3 SBD predictive analytics

The rapid growth of enterprise needs correlated with such an increase in the volume of modern data repositories on the one hand, and the nature of the data that can be stored on the other hand, have made traditional statistical methods inadequate to meet all data analysis requirements. This has necessitated the development of advanced data analytics to extract useful knowledge from such a vast volume of data.

In the light of the general perception of the advanced data analytics, a question arises about the benefits that some organisations can acquire from adopting these techniques. One of the professional sectors that has started to benefit from this notion is healthcare [76, 77]. With the increase in electronic health records, health care providers and researchers can mine the immense stores of data to detect previously unknown cognitive patterns and then use this information to build predictive models to improve diagnosis and health care outcomes.

In this context, companies incorporate advanced social data analytics to build effective marketing strategies by leveraging the interactivity enabled by online social services [78, 79]. Thus, to create the required interaction with their customers, companies use many modern means of communication to attract customers and visitors to their online social platforms. Consequently, it is necessary for companies to analyse the customers' social content and classify the customers into appropriate categories, then deliver the right message to the right category. Segmentation [80] is the first step towards effective marketing, and is intended to classify customers according to their interests, needs, geographical locations, purchasing habits, lifestyle, financial status and level of brand interaction. If companies succeed in building effective clusters of customers and determining the basic criteria for each cluster in making their buying decisions, companies will be able to establish goals and take appropriate actions to achieve them. For example, companies can identify the most optimal products/services captured for each segment of customers. This fine-grained analysis can maximise customer satisfaction as companies can then design and manufacture not only one standard product, but several segment-oriented products.

### 1.3.4 Affective design in the era of SBD

To increase the competitive of new products, product designers need to design and develop a product in order to maximize customer satisfactions. In the past, product designers only address the functional capability, reliability and efficiency of products. Nowadays, the product designers also require to address the aesthetic appreciation of the products such as outlooks, colours and shapes. This appreciation can be integrated as affective design which attempts to increase the emotional impression to the products. When the product has a certain level of affective qualities, the product can be attractively presented to consumers; the product is able to have a competitive position to the market [81].

As an example, a manufacturer is managing to develop a new car. The manufacturer requires to ensure the car efficiency and reliability, as well as multi-functional capabilities. The manufacturer mostly only attempts to ensure the basis to provide a safe, reliable, efficient and comfortable driving environment. Also, the manufacturer needs to provide multi-car functions including car cruise control, WIFI connection and audio entertainment system. To promote the car in a higher marketplace, the manufacturer also needs to satisfy the affective quality of the car. Those affective quality is correlated to the car colour, texture, shape and outlook. This is the reason why some cars in the same brand are more expensive than the others, although they have the same functional capability, reliability and efficiency. This example demonstrates that the significance of affective design, prior to manufacture a new product. The affective design has significance impact to increase the customer satisfaction of a new product.

In the past, the manufacturer needs to develop the survey questionnaires and interviews in order to collect consumer opinion of how the consumers can be satisfied with the affective quality [82]. However, developing the survey is time consuming and also significant human resource and effort is required to develop the survey questionnaires/interviews and to conduct the consumer survey. Thanks to the Internet of Things (IoT) technologies and the availability of SBD, consumer data related to affective design can be collected from product web, consumer blog and social media for consumers which discuss the affective quality and aspects of new products. Consumers review social media, consumer blog, and product web, when choosing their new products. Therefore, this SBD provides insight to the product designer to develop a product with certain affective quality. However, using the traditional data mining techniques is inefficient to analyse this SBD [18]. In this book, we will discuss the recent machine learning technologies to analyse the big data for affective design.

### 1.3.5 Social sentimental analysis

Sentiment analysis (also known as opinion mining) is the process of recognising and quantifying the emotions inferred from textual content by means of statistical analysis, natural language processing, computational linguistics etc. [83]. In the last few years, social media has been successfully utilising some of the research done in the area of social sentiment analysis. The importance of sentiment analysis in the social media context comes from its utility for market analysis, listening to the voice of customers and to feed business intelligence applications with harvested customers' feelings toward a particular product or service [84]. This facilitates businesses with the ability to provide a better and rapid customer service. Further, social sentiment analysis can be used in various other applications, such as spam detection [85], stock movement prediction [86], disaster relief [87], social credibility [88], and many other applications. Therefore, social sentiment analysis has become a core dimension of researchers' endeavours to create applications that leverage the massive increase in user-generated content as well as utilise the advances in big data technologies [89].

In this book, we will discuss the arena of sentiment analysis in the context of SBD. In particular, the book will depict approaches that incorporate Big Data technologies to track sentiments and opinions captured from public social media. The book will interpret various technical terminology related to sentiment analysis. Also, we discuss an incorporated big data research framework that includes various approaches and then deliberate on the models utilized in our research, including the experimental design employed.

## 1.4 Conclusion

Considerable achievements have been made in SBD analytics motivated by the need for efficient and effective social data analytics solutions. This chapter introduces the notion of SBD and presents several applications and analytics that are beneficial for SBD analytics. In particular, various definitions of SBD terms are presented and a selected set of social media services are discussed. The chapter also attempts to create concrete ground on interconnected important applications that can be carried out on SBD. Next chapter will provide an overall depiction to the notion of Big data followed by a discussion on characteristics of Big data and the incorporated technology that is commonly used in industry and academia.

18......